\documentclass[11pt,titlepage]{amsart}
 \usepackage{amsaddr}
 
\usepackage{graphicx}        
\usepackage{multicol}        
\usepackage[bottom]{footmisc}

\usepackage{amsmath}
\usepackage{amssymb}
\usepackage{morefloats}
\usepackage{xcolor}
\usepackage{rotating}
\usepackage{caption}
\usepackage{xspace}
\usepackage{algorithm}
\usepackage{algorithmic}

\def\LG{\textsl{Lurker Game}\xspace}

\begin{document}

\title{Modeling Evolutionary Dynamics of Lurking in Social Networks}

\author{Marco A. Javarone}
\address{Dept. of Mathematics and Computer Science, University of Cagliari, Italy}
\email{marcojavarone@gmail.com}

\author{Roberto Interdonato}
\address{Dept. of Computer, Modeling, Electronics, and Systems Engineering, University of Calabria, Italy}  \email{rinterdonato@dimes.unical.it} 

\author{Andrea Tagarelli}
\address{Dept. of Computer, Modeling, Electronics, and Systems Engineering, University of Calabria, Italy} \email{andrea.tagarelli@unical.it}

\maketitle

\begin{abstract}
Lurking is a complex user-behavioral phenomenon that occurs  in all large-scale online communities and social networks. It generally refers to the  behavior characterizing  users that benefit from the information produced by others in the community without actively contributing back to the production of social content.   
The amount and evolution   of lurkers may strongly affect an online social environment, therefore understanding the lurking dynamics     and identifying strategies to curb this trend are relevant problems. 
In this regard, we introduce the \textit{Lurker Game}, i.e., a model for analyzing the transitions from a lurking to a non-lurking (i.e., active) user role, and vice versa,    in terms of evolutionary game theory.  
We evaluate  the proposed Lurker Game by arranging agents on complex networks and analyzing  the system evolution, seeking relations between the network topology and the final equilibrium of the game.  
Results suggest that the Lurker Game is suitable to model the lurking dynamics, showing    how   the adoption of rewarding  mechanisms combined with the modeling of hypothetical heterogeneity of users' interests  may lead users in an online community   towards a cooperative behavior.  
\end{abstract}

\section{Introduction}\label{sec:Introduction}
Most members of online communities and social networks do not actively contribute to the shared online space, i.e., they only consume (e.g., read, watch) information without sharing their knowledge or expressing their opinion. These users are commonly defined as \emph{lurkers}, since they remain quite unnoticed while benefiting from others' information or services.  
Remarkably, lurkers feel themselves as community members, and should not be trivially regarded as totally inactive users, i.e., registered users who do not use their account to join the online community.   

The characterization of lurking in online communities has been a controversial issue  from a social science and computer-human interaction perspective~\cite{Edelmann13}.  
One common perception of lurking is related to the infrequency of active participation to the community life~\cite{PreeceNA04}, 
while other definitions refer to legitimate peripheral participation~\cite{LaveW91},  
individual information strategy of microlearning~\cite{Kahnwald2006},  and knowledge sharing barriers~\cite{Ardichvili08}.  
In general, in the realm of online social networks (OSNs), neutral or even positive views of the presence of lurkers have normally supplanted negative views. The silent presence of lurkers can indeed be seen as harmless as it  reflects a subjective reticence (rather than malicious motivations) to contribute to the community wisdom~\cite{PreeceNA04}. Moreover, lurking can be expected or even encouraged because it allows  newcomers to learn the netiquette before they might decide to provide a valuable contribution over time. On the other hand, if users are worried that their private information may be revealed or their security may be threatened by posting, they may decide to lurk to protect themselves.

Lurkers hold great  potential in terms of \textit{social capital}, because they acquire knowledge from the OSN; further, they might decide to use this knowledge in order to form their own opinions, although these will never or rarely be unveiled to the community. Within this view, it is highly desirable to \textit{delurk} such users, i.e., to apply a mix of strategies aimed at encouraging lurkers to return their acquired social capital, through a more active participation to the community life.     
As a matter of fact, even though a massive presence of lurkers is typical in a   large-scale social environment, too many lurkers  would impair the virality of the online community, which    instead needs to be sustained over time with fresh ideas and initiatives.    
Social science and human-computer interaction research studies have addressed the delurking problem mainly focusing on the conceptualization of the strategies to adopt, such as~\cite{Sun+14}: reward-based external stimuli (e.g., badges~\cite{leskovec01}), providing encouragement information, improvement of the usability of the online platform, and  guidance from elders/master users to help lurkers become familiar with the system as quickly as possible. 
However, given the variety of influencing factors that drive online participation, developing a  computational approach to turn lurkers into active members of an OSN is an emerging yet challenging problem, regardless of the delurking strategy adopted. 

\ 

\textbf{Contributions.\ } 
Our intuition in this work is that the behavioral dynamics underlying the transition from a lurking to  non-lurking (i.e., active) user role, and vice versa, can suitably be modeled via an \textit{evolutionary game  theory} approach~\cite{moreno03,tomassini01,moreno02,perc03,javarone01}.  
%
We define the \LG, in which active users are regarded as \textit{cooperators} and lurkers as \textit{defectors}. Cooperators contribute to the system by adding information represented by ``virtual coins'' to a common pool, while defectors do not contribute. The total amount of virtual coins in the common pool increases according to two key aspects: (i) the collective effort of cooperators and (ii) the different impact that information naturally has on each agent, depending on her/his preferences.
Our \LG employs a Fermi-like function~\cite{perc01} to model the transition probability from one agent strategy to another. 
Having considered the importance of rewarding mechanisms~\cite{perc01,perc02,leskovec02} towards an ordered phase of cooperation (i.e., delurking), we also introduce a prize structure for promoting cooperation.
We evaluate the \LG on random graph models that resemble the complexity of real-world OSNs, focusing on the effect that the network topology may have on the final equilibrium of the game. 
This work represents, to the best of our knowledge,  the first attempt for quantitatively understanding lurking and delurking dynamics in OSNs  via the evolutionary game theory.

The remainder of the paper is organized as follows.  Section~2  introduces the \LG on complex networks. Section~\ref{sec:results} shows results of numerical simulations, and Section~\ref{sec:discussion} provides a discussion on main experimental findings.   
Related works are discussed in Section~\ref{sec:related}, finally Section~\ref{sec:conclusions} concludes  the paper. 

\section{The \LG}\label{sec:model}
User-generated communications and social content produced in an OSN represent a rich source of knowledge whose value can, in principle, be increased by collective efforts. Within this view, evolutionary games provide a powerful tool to model the dynamics of  OSNs~\cite{abramson01,segbroeck01,nowak02}. 

Our aim in this work is the definition of a novel game, named \LG, to analyze the dynamics of OSN populations, by focusing on the two main roles played by network members: active contributors and lurkers.  The former are regarded as   contributors, whereas the latter as  defectors. 
Information generated by contributors is expressed in terms of \textit{virtual coin} ($vc$), which is assumed to be unitary by default. Note that we adopt the term information with its more general meaning, which includes any type of social content produced in an OSN (i.e., posts, comments, preferences, etc.). 

Our \LG entails important aspects in cooperator-defector games. 
The collective effort is represented by a synergy factor $r$ ($r>0$), which is usually  adopted in \textit{public goods games} (PGGs)~\cite{perc01,perc02}, and used to grant groups of cooperators. 
However, \LG has two main differences  from classic PGG. First,  due to its nature, the ``public goods'' in our game (i.e., information generated by contributors)  is not divided but rather \textit{equally shared} among all users of a group. Second,    we observe that information may acquire a different value for each individual (e.g.,   one may contribute by writing posts on politics but it is not interested in reading about music); to model this heterogeneity of user interests and preferences, we introduce a further parameter, denoted by $\nu$, ranging in  $(0,1]$, such that  the common pool of virtual coins, shared in the OSN environment, is diversified by means of $\nu$.

\subsection{Basic Dynamics}
Given a set of $N$ agents, the dynamics of \LG unfolds in discrete time steps and is defined  as follows. At each time step, agents have to put into the common pool a virtual coin if they take the role of cooperators, otherwise (i.e., they are lurkers) do nothing. The accumulated amount of virtual coins is increased by $r$ and $\nu$, and then equally shared among all agents. The \textit{payoff equations} in \LG are defined as follows: 
\begin{equation}\label{eq:lg_payoff}
\begin{cases}
\pi^{c} = r \nu \sum_{1}^{N^c} vc - vc\\
\pi^{d} = r \nu \sum_{1}^{N^c} vc
\end{cases}
\end{equation}
with $N^c$ number of cooperators, $r$ synergy factor, and $\nu$ representing the heterogeneity of interests of users. 
Due to its evolutionary nature, \LG allows agents to change their strategy~\cite{tomassini01}, i.e., from cooperation to defection and vice versa. In particular, when considering two agents at a time,  we adopt a Fermi-like function to implement a transition probability from one strategy to another. Given two agents  $x$ and $y$, this probability is defined as:  
\begin{equation}\label{eq:fermi_function}
W(s^x \to s^y) = \left(1 + \exp\left[\frac{\pi^y - \pi^x}{K}\right]\right)^{-1}
\end{equation}
where $s^x$ and $s^y$ denote the strategies of the players $x$ and $y$, respectively, $\pi^x$ and $\pi^y$ denote their respective payoff, and $K$ indicates uncertainty in adopting a strategy. By setting $K = 0.5$, we implement a rational and meritocratic approach during the strategy revision phase~\cite{perc01}.
Like in the PGG, behaving as defectors is much more convenient than behaving as cooperators and the Nash equilibrium of \LG corresponds to defection.

\subsection{Mean Field Analysis}
We perform a  \textit{mean field} analysis~\cite{barra01} of \LG, in order to investigate if the Nash equilibrium corresponds to the final ordered phase. 
Hence, we assume that the population is composed of only one big community and every agent interacts with all the others.
Under this assumption, the evolution of a population with $N$ agents is described by the following set of equations~\cite{javarone04}:
\begin{equation}\label{eq:lotka_volterra}
\begin{cases}\frac{d\rho^c(t)}{dt} = p^c \cdot \rho^c(t) \cdot \rho^d(t) - p^d \cdot \rho^d(t) \cdot \rho^c(t)\\ 
\frac{d\rho^d(t)}{dt} =  p^d  \cdot \rho^d(t) \cdot \rho^c(t) - p^c\cdot \rho^c(t) \cdot \rho^d(t)\\
\rho^c(t) + \rho^d(t) = 1
\end{cases}
\end{equation}
\noindent with $\rho^c(t)$ and $\rho^d(t)$ densities of cooperators and defectors, $p^c(t)$ probability that cooperators prevail, and $p^d(t)$ probability that defectors prevail. These probabilities are computed according to the payoffs obtained, at each time step, by cooperators and defectors as defined in Eq.~\ref{eq:lg_payoff}.
Therefore, we have to consider the difference between the payoffs accumulated by the two   agents  randomly chosen at each time step.
If we denote with $x$ a cooperator and with $y$ a defector, the probability  $p^c$ corresponds to $W(x \to y)$, so we consider the difference $\pi^{(d)} - \pi^{(c)}$. While, $p^d$ corresponds to $W(y \to x)$, then we consider $\pi^{c} - \pi^{d}$. 
Few algebraic steps lead to the following solutions:
\begin{equation}\label{eq:payoff-diff}
\begin{cases}
\pi^{d} - \pi^{c} = r \nu  N \rho^c - r \nu  N \rho^c + 1 = 1\\ 
\pi^{c} - \pi^{d} = r \nu  N \rho^c - 1 - r\nu  N \rho^c  = -1
\end{cases}
\end{equation}
By substituting results of Eq.~\ref{eq:payoff-diff} in Eq.~\ref{eq:fermi_function}, one obtains $p^c \sim 0.12$ and $p^d \sim 0.88$. Remarkably, the mean field approach to \LG leads to dynamics completely independent both from $r$ and $\nu$. 
Given the values computed in Eq.~\ref{eq:payoff-diff}, the solution of the system in Eq.~\ref{eq:lotka_volterra}   confirms the expected result, i.e., defection prevails according to the Nash equilibrium.

\subsection{Rewarding Mechanisms}
The above result leads us to focus on \textit{rewarding mechanisms} to drive a population towards an ordered phase of cooperation. 
%
Therefore, we introduce a   variation in the basic formulation of \LG by introducing a prize structure for promoting cooperation. This impacts on the payoff equation of cooperators, whereby  the set of payoff equations is modified as follows:
\begin{equation}\label{eq:lg_payoff_badge}
\begin{cases}
\pi^{c} = r \nu \sum_{1}^{N^c} vc - vc + \Phi(\Delta t^c)\\
\pi^{d} = r \nu \sum_{1}^{N^c} vc
\end{cases}
\end{equation}
\noindent with $\Phi(\Delta t^c)$ rewarding function that allows cooperators to receive a further amount of virtual coins. This function takes in input $\Delta t^c$, i.e., the amount of time each agent behaves as a cooperator.
The prize structure $S$ grants cooperative agents at a fixed rate, i.e., every $k$ time steps: $S: \Delta t^c = \{k, k, \ldots, k\}$.
This way, each prize consists of an amount of $vc$ equal to that paid by a cooperator over time (between two achieved prizes). We define the \textit{prize function} as follows:
\begin{equation}\label{eq:lg_payoff}
\Phi(\Delta t^c) = \begin{cases}
 \Delta t^c \cdot vc  & \mbox{if } \Delta t^c \in S\\
0 & \mbox{if } \Delta t^c \not \in S
\end{cases}
\end{equation}
Analogously to the basic dynamics of \LG, after every iteration agents undergo a strategy revision phase based on Eq.~\ref{eq:fermi_function}. 
Algorithm \ref{algo:LG} sketches the main steps performed in \LG.

%
%

\begin{algorithm}[t!]
\caption{ \LG  \label{algo:LG}}
\begin{algorithmic}[1]
\REQUIRE A population of $N$ agents, where $N^c$ are cooperators and $N^d$ are defectors ($N=N^c+N^d$). \\ The synergy factor $r >0$. \\ The user preference coefficient $\nu \in (0,1]$. \\
A network topology  $G$ that models the connectivity of the $N$ agents, otherwise agents are fully connected to each other (mean field). 
\STATE \textbf{repeat} 
\STATE \quad Compute the payoff of cooperators and defectors, according to Eq.~\ref{eq:lg_payoff_badge}
\STATE \quad Randomly select two agents $x$ and $y$ (with different strategies) s.t. $x,y$ are linked w.r.t. $G$
\STATE \quad Agent $y$ takes the strategy of agent $x$ according to Eq.~\ref{eq:fermi_function}
\STATE \textbf{until} all agents have the same behavior (Nash equilibrium)
\end{algorithmic}
\end{algorithm}

\subsection{\LG on Networks}
Since social networks constitute the natural environment to observe the phenomenon of lurking, we also study the \LG on complex networks. 
Following the lead of previous studies on evolutionary games  (e.g.,~\cite{nowak01,nowak02,szabo01,moreno03,perc05,javarone01}), we focus our attention on two relevant models: Barabasi-Albert model~\cite{barabasi01} (hereinafter BA) and Watts-Strogatz~\cite{watts01} model (hereinafter WS).
Since the topological properties of networks generated by both considered models (i.e., BA and WS) are well-known (see, e.g.,~\cite{barabasi02}), all outcomes of the proposed model can be analyzed seeking relations with the considered topology.

Note that when agents are arranged on networks, the dynamics of the game are different from those adopted in the mean field case, which in topological terms, can be viewed as a fully-connected network. Adopting complex networks, only few agents are considered at each iteration. In particular, at each time step two randomly chosen agents play \LG with all groups of belonging.
Therefore, the accumulated payoffs are computed for each group and the final prize is assigned only to cooperative agents that played the game.
Next, as previously discussed, the $x$-th agent tries to enforce its strategy to the $y$-th agent with probability defined in Eq.~\ref{eq:fermi_function}.

\textbf{Memoryless and memory-aware payoff.\ } 
We introduce a further aspect of the proposed model, related to the way agents manage their accumulated payoffs. We distinguish between two scenarios of payoff accumulation, namely  \textit{memoryless} and \textit{memory-aware}.

The memoryless case entails that every time two agents are selected to play \LG with their groups, they reset their accumulated payoff. Therefore, when computing the transition probability of Eq.~\ref{eq:fermi_function}, they consider only the payoff accumulated during the present time step. Instead, the memory-aware case entails agents save their payoff over time. 
Note that while the memory-aware case is closer to a real scenario (e.g., online users may accumulate several badges over time), the memoryless case avoids noise effects in numerical simulations that can emerge in Eq.~\ref{eq:fermi_function} for large payoffs.

We investigate both cases, by introducing a \textit{cutoff} in the difference between the payoffs of the two considered agents (i.e., $x$ and $y$). In doing so, for large payoffs, the Fermi function behaves like a simple rule with only two possible results: $1$ and $0$, i.e., $1$ if the payoff of the $x$-th agent is greater than that of the $y$-th, and $0$ otherwise. Thus, the interesting granularity, in terms of transition probabilities, introduced by the Fermi function is lost in the memory-aware case, after few time steps.

It is also relevant to observe that a similar problem may arise when dealing with scale-free networks since, even in the memoryless case, nodes with high degree (i.e., hubs) can accumulate at each iteration a very high payoff.
As a result, we expect that simulations performed on scale-free networks in the memoryless case yield outcomes similar to those achieved by the memory-aware case, at least by considering the same topology (i.e., scale-free in both cases).

\textbf{Identifying critical parameters}
Numerical simulations will be primarily devoted to the identification of critical values of $k$ and $\nu$, i.e., the step adopted in the prize structure $S$ and the variety of information (or users' interests) in the social network, respectively. These values   together with the final equilibrium achieved in both networks, provide a useful indicator for studying the dynamics of \LG and for comparing different network topologies.
Remarkably, we are dealing with a disordered system~\cite{galam01,javarone03,javarone04}, in terms of states (i.e., cooperators and defectors), having only two possible equilibria: one characterized by the prevalence of one species (i.e., cooperators or defectors) and one characterized by a coexistence of both species at equilibrium. The former corresponds to a ferromagnetic phase, whereas the latter to a paramagnetic phase~\cite{javarone03}.
Thus, both the Nash equilibrium  and its opposite case  correspond to the ferromagnetic phase. The paramagnetic phase has been observed in games like the PGG, obtained by tuning the synergy factor and without adopting rewarding mechanisms~\cite{perc01}. 
%

\section{Results}  \label{sec:results}
\begin{figure}[t!]
\centering
\begin{tabular}{ccc}
\includegraphics[width=0.33\textwidth]{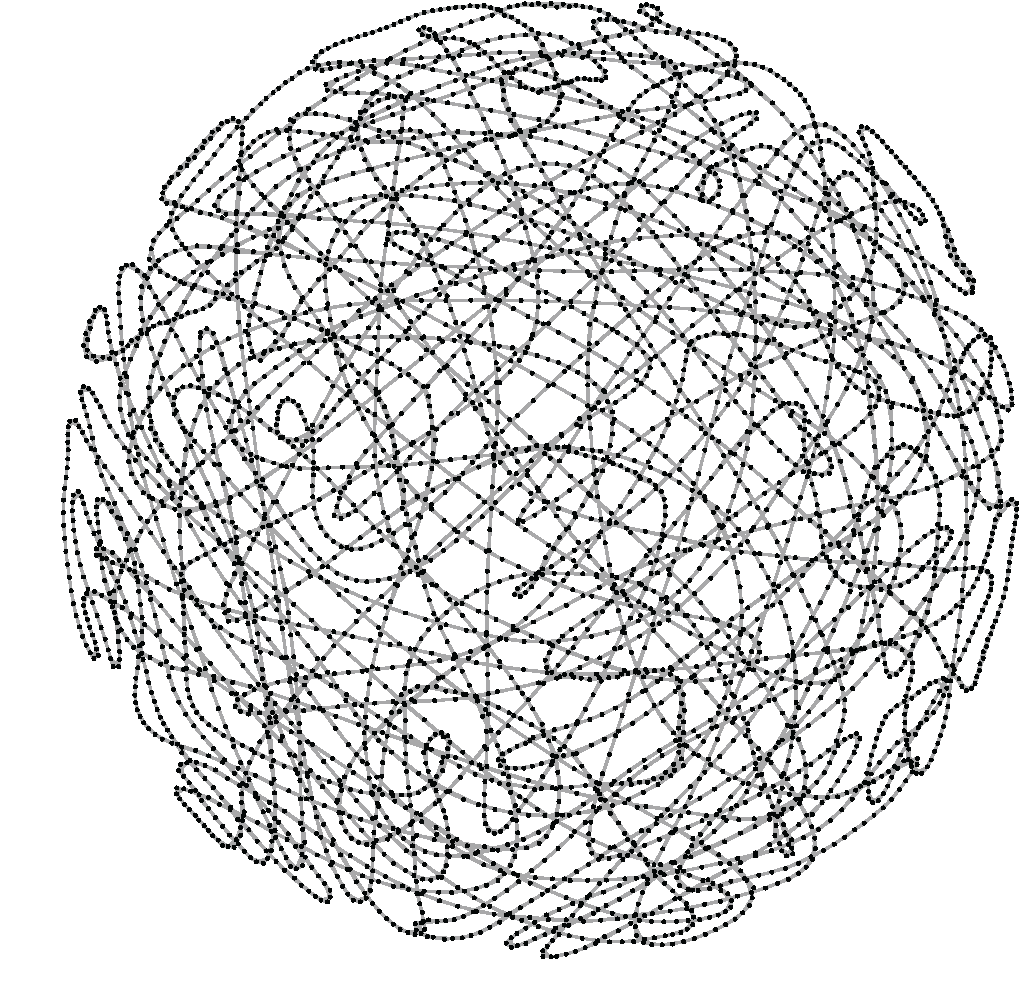} & 
\includegraphics[width=0.33\textwidth]{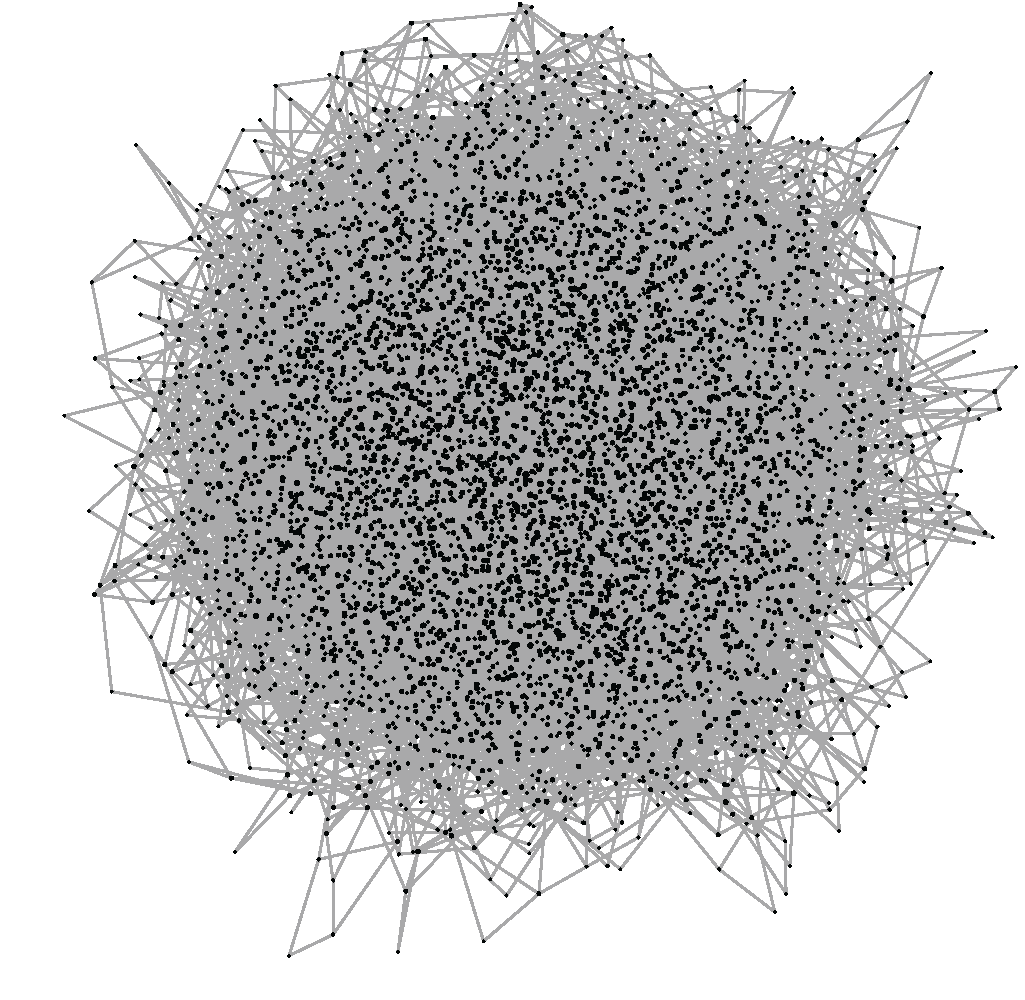} & 
\includegraphics[width=0.33\textwidth]{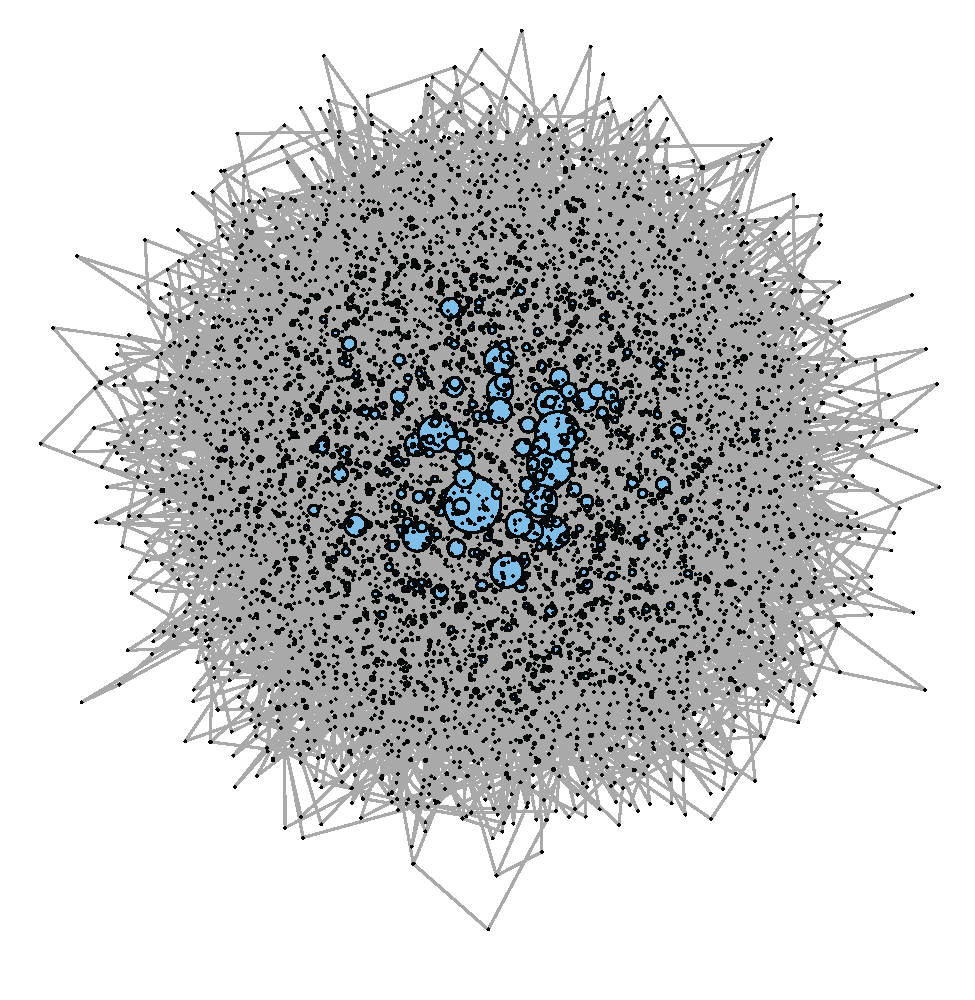} \\
(a) & (b) & (c) 
\end{tabular}
\caption{Evaluation networks: 
(a) WS with $\beta=0.0$, (b) WS with $\beta=0.5$, (c) BA.}
\label{fig:drawings}
\end{figure}

\textbf{Experimental setting.\ } 
We evaluated \LG by arranging agents on different networks, generated according to the BA and WS models. 
The former generates scale-free networks, i.e., networks characterized by the presence of nodes with a very high degree, defined hubs.
The WS model generates different kinds of networks by tuning a rewiring parameter, $\beta$, which ranges within $[0,1]$. In particular, $\beta = 0$ yields a regular ring lattice topology,   intermediate values of $\beta$ yield small-world-networks (characterized by relatively low  average path lengths and  high clustering coefficients), while completely random networks   are  obtained for high values of $\beta$.  
In this work, we considered the following values: $\beta = \{0.0, 0.3, 0.5, 0.8\}$.
Figure~\ref{fig:drawings} shows a pictorial representation of each kind of networks, whereas Table~\ref{tab:data} reports some of their structural properties (achieved with $5000$ nodes). 

\begin{table}[t!]
\caption{Structural properties of evaluation networks.}
\label{tab:data}
\centering
\begin{tabular}{lcccc}
\hline
\textit{network model}~~~&~~~\textit{avg. path length}~~~&~~~\textit{diameter}~~~&~~~\textit{clust. coeff.} \\
\hline\hline
WS $\beta=0.0$   &  625.38 & 1250 & 0.500 \\ 
WS $\beta=0.3$   & 7.89 & 14 & 0.165 \\
WS $\beta=0.5$   & 6.99 & 12 & 0.054 \\
WS $\beta=0.8$   &  6.67 & 11 & 0.005 \\
BA  &   4.85 & 9 & 0.002 \\
\hline
\end{tabular}
\end{table}

Numerical simulations were performed with $N = 5000$ agents, with an equal initial density of cooperators and defectors (i.e., $\rho^c(0) = \rho^d(0) = 0.5)$, and an average degree $\langle k\rangle = 4$.
We set the synergy factor $r$ to $2$, as we found that this value does not allow cooperators to survive without rewarding mechanisms (see also~\cite{perc01} for a discussion about the critical thresholds of the synergy factor). Parameter $\nu$ was instead varied considering values from $0$ to $1$. It is worth noting  that for $\nu = 0.2$ the game, in the memoryless case and without the adoption of rewarding mechanisms, corresponds to the PGG in networks with the same topology.
Simulations were carried out for a maximum number of time steps equal to   $10^8$, then results were averaged over several different runs.\\

\textbf{Evolution of the system.\ } 
We initially analyzed the density of cooperators over time in all networks.
\begin{figure}[t!]
\centering
\begin{tabular}{cc}
\includegraphics[width=0.5\textwidth]{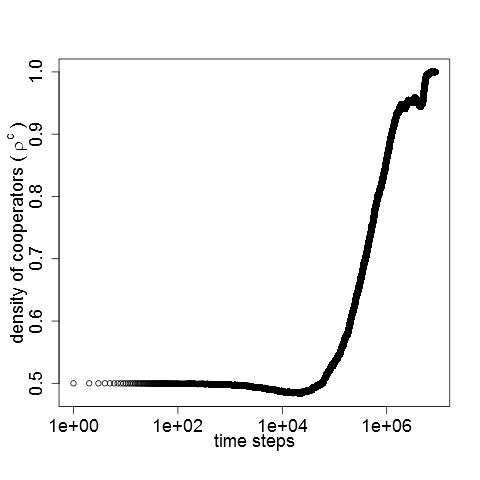} & 
\includegraphics[width=0.5\textwidth]{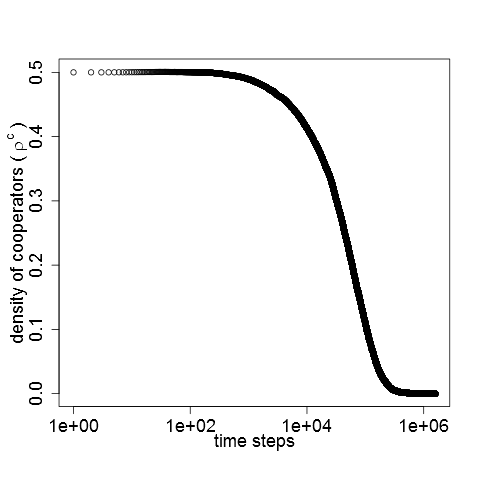} \\
(a) & (b) 
\end{tabular}
\caption{Possible behaviors of the  \LG system. Time evolution  of density of cooperators: (a) cooperators prevail, (b) defectors prevail. Results correspond  to WS model ($\beta = 0.5$) with 5000 agents and $k=2$, for $\nu$ equal to (a) $0.5$ and (b) $0.3$.}
\label{fig:figure_1}
\end{figure}
We found three  main behaviors: cooperators vanish (Fig.~\ref{fig:figure_1}(b)) or prevail (Fig.~\ref{fig:figure_1}(a)) after a number of time steps, or both cooperators and defectors coexist over time.
This finding clearly indicates that a population playing the \LG can reach both ordered phases and disordered phases at equilibrium.
In particular, since agent strategies can be mapped to spins $\sigma \pm 1$ respectively and, as observed, there are only two possible equilibria, the evolution of the system can be analyzed in terms of ferromagnetic phase transitions~\cite{mobilia01,barra01}. 
%
%
Thus, mapping our model to a spin system allows us to identify the conditions that can lead towards the different kinds of equilibrium. The relevance of identifying a description based on the language of phase transitions, lays in the fact that it opens the way to further analytical investigations~\cite{javarone04} that can potentially lead to get new insights on the proposed model.

\textbf{Critical values of $\nu$.\ } 
We finally analyzed the role of $\nu$. Results are reported in Table~\ref{tab:memoryless} for the memoryless case. For the memory-aware case, results indicate a more complex scenario, which is discussed next.  
\begin{table}[t!]
\caption{Memoryless Agents}
\centering 
%
\begin{tabular}{ccc}
\hline
\textit{network model}~~&~~Critical $\nu$~~&~~$k$-range\\
\hline \hline
WS $\beta=0.0$ & $0.6$ & $[1..5]$ \\
WS $\beta=0.3$  & $0.44$ & $[1..5]$\\
WS $\beta=0.5$  & $0.42$ & $[1..5]$\\
WS $\beta=0.8$ &$0.41$ & $[1..5]$\\
\hline
BA  & $0.22$  & $[1..5]$\\
 \hline
\end{tabular}
\label{tab:memoryless}
\end{table}

\section{Discussion} \label{sec:discussion}

Results of our investigations suggest that our \LG has      a rich behavior, which can be described by considering the main degrees of freedom of the system: $\nu$, $k$, network topology and the evolution of payoffs over time.

\textbf{Results on WS networks.\ } 
In the memoryless case, for each considered $\beta$, we found a well recognized critical $\nu$. In particular, by increasing $\beta$, cooperators require a smaller $\nu$ to prevail. This   suggests that, in general, random topologies support cooperation better than regular ones.
It is worth noting that in all cases critical $\nu$ showed a certain robustness towards the considered $k$ values, i.e., $k \in [1..5]$. In this regard, further investigations will be devoted to better clarify the relation between $\nu$ and $k$, since we hypothesize that for high $k$ values defectors may prevail even for $\nu$ values greater than the identified thresholds (see Table~\ref{tab:memoryless}).
On the other hand, results achieved by memory-aware agents indicate that, in general, critical $\nu$ are smaller than those found in the memory-less case, e.g., for $\beta  = 0.0$ we obtained $\nu \sim 0.4$. %
However, we found that even for values greater than the minimal threshold of $\nu$, sometimes defectors may prevail. Before trying to mind a hypothesis about this behavior, we have to recall that in the memory-aware case some noise may arise resulting from high payoffs. Moreover, the memory-aware case may easily promote cooperation than its counterpart as groups of cooperative agents tend to increase their payoff unboundedly. Therefore, as a future work, we aim to investigate this   aspect of the model.
Also note that  a mixed phase (i.e., composed of both species) has been found for values close to the critical $\nu$.

\textbf{Results on scale-free networks.\ } 
When considering the BA model, a major finding is that cooperators need a smaller $\nu$ to prevail than those computed in WS network; specifically,  $\nu = 0.22$ and $\nu \sim 0.1$, in the memory-less and in the memory-aware case, respectively.  
Moreover, scale-free networks in the memory-aware case show an interesting bistable behavior for small values of $\nu$. We suggest again that this may result from noise introduced by the utilization of large payoff in the Fermi function that we faced by adding a numerical cutoff. 
It is relevant to note that our results are in accord with those reported  in~\cite{santos01}, as stated above, since scale-free networks have been found to foster cooperation better than other topologies. Also, like   for   WS networks, critical $\nu$ are robust to variations of $k$ in the considered range. 

Overall, the proposed \LG suggests that   the adoption of rewarding  mechanisms combined with the modeling of hypothetical heterogeneity of users' interests ($\nu$)   may lead a population towards a cooperative behavior. 
This supports  our initial intuition that \LG is suitable to model the dynamics of such a  complex phenomenon  as lurking. 

\section{Related work}  \label{sec:related}
In~\cite{ASONAM,SNAM14}, the authors developed the first  computational approach to lurker mining, focusing on ranking problems. To this purpose, they proposed a topology-driven definition of lurking behavior, based  on  principles of   overconsumption, authoritativeness of the information received, and non-authoritativeness of the information produced. Quantitative and qualitative evaluation results showed how the proposed methods are effective in identifying and ranking lurkers in real-world OSNs.

The same authors also posed a first step toward the definition of delurking strategies in~\cite{InterdonatoPT15}, by proposing a  targeted influence maximization problem under the linear-threshold diffusion model. In this context, a set of previously identified lurkers  is taken as target set of an influence maximization problem, whose  
objective function   is defined   upon the concept of \textit{delurking capital}, i.e., the social capital gained by activating lurkers in an online community. 

We can also mention   research studies that, though not specifically concerning lurking, addressed related problems in OSNs via a game-theoretic approach. 
For instance, Anand et al.~\cite{AnandCSV13} defined a Stackelberg game  to maximize the benefit each user gains extending help to other users, hence to determine the advantages of being  altruistic. Some interesting remarks relate the altruism of users to their level of capabilities, and indicate that  the benefit derived from  being altruistic  is larger than that reaped by selfish users or free riders.  
Malliaros and  Vazirgiannis~\cite{MalliarosV13} also built  upon game theory to study the property of users' departure dynamics, i.e.,  the tendency of individuals to leave the community.  
 
 Our proposed approach in this work differs from all the aforementioned studies as it represents both a novel computational approach to lurking and delurking user-behaviors, and a novel application domain in the field of evolutionary games.

\section{Conclusion} \label{sec:conclusions} 

In this work, we brought for the first time evolutionary game theory into the analysis of lurking behaviors in OSNs. 
We defined the \LG and evaluated it through both a mean-field analysis and     by arranging agents on small-world and scale-free networks. Results suggest that \LG is suitable to model the dynamics of such a  complex phenomenon  as lurking, showing   a rich behavior depending on the network topology and on the way agents manage their payoff. Remarkably,  \LG allows us to understand  how   the adoption of rewarding  mechanisms combined with the modeling of hypothetical heterogeneity of users' interests  may lead a population towards a cooperative behavior.    
Further investigations will be mainly devoted to better clarify the interrelation between the two model parameters in \LG, also including analysis over other network topologies and larger populations.

\end{document}